\documentclass[onecolumn,preprintnumbers,amsmath,amssymb,nofootinbib]{revtex4}

\usepackage{graphicx}
\usepackage{bm}
\usepackage{amsmath}

\newcommand{\be}{\begin{equation}}
\newcommand{\ee}{\end{equation}}
\newcommand{\bea}{\begin{eqnarray}}
\newcommand{\eea}{\end{eqnarray}}
\newcommand{\bdm}{\begin{displaymath}}
\newcommand{\edm}{\end{displaymath}}
\newcommand{\beas}{\begin{eqnarray*}}
\newcommand{\eeas}{\end{eqnarray*}}

\newcommand{\tphi}{\tilde{\phi}}
\newcommand{\tpsi}{{\psi}}
\newcommand{\tPhi}{\tilde{\Phi}}
\newcommand{\tPsi}{\tilde{\Psi}}
\newcommand{\tB}{\tilde{B}}
\newcommand{\tE}{\tilde{E}}
\newcommand{\tg}{\tilde{g}}
\newcommand{\tG}{\tilde{G}}
\newcommand{\tV}{\tilde{V}}
\newcommand{\tT}{\tilde{T}}
\newcommand{\tNabla}{\tilde{\nabla}}
\newcommand{\tR}{\tilde{R}}
\newcommand{\ta}{\tilde{a}}
\newcommand{\trho}{\tilde{\rho}}
\newcommand{\tP}{\tilde{P}}

\newcommand{\tH}{\tilde{\mathcal{H}}}
\newcommand{\tdelta}{\tilde{\delta}}

\begin{document}
\title{Gauge Issues in Extended Gravity and $f(R)$ Cosmology}
\author{Iain A. Brown}
\email{ibrown@astro.uio.no}
\author{Amir Hammami}
\email{amirham@astro.uio.no}
\affiliation{Institute of Theoretical Astrophysics, University of Oslo, P.O. Box 1029 Blindern, N-0315 Oslo, Norway}

\begin{abstract}
We consider issues related to the conformal mapping between the Einstein and Jordan frames in $f(R)$ cosmology. We consider the impact of the conformal transformation on the gauge of a perturbed system and show that unless the system is written in a restricted set of gauges the mapping could produce an inconsistent result in the target frame. Newtonian gauge lies within the restricted group but synchronous gauge does not. If this is not treated carefully it could in principle contaminate numerical calculations.
\end{abstract}

\maketitle

\section{Introduction}
\noindent Extended gravity theories, where the Einstein-Hilbert Lagrangian density $\mathcal{L}_{\rm EH}=R$ is replaced by a more general function including terms of higher-order in derivatives of the metric ($R^2$, $R_{\mu\nu}R^{\mu\nu}$, $R_{\alpha\beta\mu\nu}R^{\alpha\beta\mu\nu}\ldots$) and couplings to new dynamical degrees of freedom, have long been of interest in relativity. The last few decades have seen increasing applications of these models to cosmology. See \cite{Capozziello:2011et,Nojiri:2006ri} and their references for an overview and further details on these models and their applications to cosmology.\footnote{Recently \cite{Biswas:2011ar} presented the most general action guaranteed to be free of ghosts when quantising perturbations on a classical vacuum; a viable model of extended gravity will be constrained to conform with this action.} In the last decade attention has focused on exploiting extended gravity to model dark energy without the need to introduce exotic particle species. A model frequently employed in this context is $f(R)$ gravity (see for example \cite{DeFelice:2010aj,Nojiri:2010wj} for recent reviews), where the Einstein-Hilbert Lagrangian density is replaced with an arbitrary function of the Ricci scalar, $\mathcal{L}=f(R)$, while the matter couples minimally to the metric and follows its geodesics in free motion. This representation of an extended gravity model is known as the ``Jordan frame''.

The action can be transformed to a variety of different forms. One of the most 
common transformations is into the so-called ``Einstein frame'' where the action is manipulated to isolate a Ricci scalar of a new metric. The residual terms can be interpreted as an effective scalar field to which matter is non-minimally coupled and deflected from the geodesics of the metric. The field equations are otherwise those of standard general relativity. We review the model in the Jordan and Einstein frames in \S\ref{Section:Model}.

Transforming from the Jordan frame to the Einstein frame is extremely useful in the study of $f(R)$ gravity. We employ the metric formulation, in which the action is varied with respect to the metric alone, and in which the field equations are fourth-order. In the Einstein frame, as in standard GR, the theory is second-order -- a significant simplification. Aspects of the transformation have been controversial for some time (see for example \cite{Faraoni:1998qx,Deruelle:2010ht,Quiros:2011wb,Capozziello:2011et} and further references in \S\ref{Section:Equivalence}). The discussion has centred upon the nature of the equivalence, and authors can be separated \cite{Faraoni:1998qx,Capozziello:2011et} into two camps: those who feel the equivalence is ``physical'' and observables can be calculated in either frame, and those who feel the equivalence is mathematical in nature and that observables should be calculated in a chosen ``physical frame''. We briefly discuss this issue in \S\ref{Section:Equivalence}.

Modern cosmology is the study of Friedmann-Lema\^itre-Robertson-Walker (FLRW) spacetimes, perturbed to linear or higher orders, and this provides an illustration of the major motivation for our paper. Numerical studies targeting the cosmic microwave background (CMB) and the matter power spectrum (such as \cite{Bean:2006up,MGCAMB11,Gu:2011kk,Li:2011vk,Dossett:2011tn}) often employ modifications to Boltzmann codes such as CAMB \cite{CAMB}. While it is common to employ a general parameterisation \cite{MGCAMB11,Gu:2011kk,Li:2011vk,Dossett:2011tn} in the Jordan frame, working in the Einstein frame (e.g. \cite{Schimd:2004nq,Bean:2006up}) provides a flexible alternative. Exploiting the Einstein frame allows us to consider any modified gravity model which possesses an Einstein frame and for which background solutions in the Jordan frame might be extremely difficult to find, either analytically or numerically. A recent example is the study of the string gas cosmology (see e.g. \cite{Brandenberger:2011et}), in which the primordial power spectrum is frequently calculated in the Einstein frame as opposed to the string frame, as in \cite{Brandenberger:2006pr,Kaloper:2006xw}. The use of the Einstein frame allows us to import our intuition and understanding from standard gravity -- at least while we remain within the frame -- and, perhaps more importantly, it allows us to directly leverage well-tested codes developed for general relativity and which contain a wealth of physics and arbitrary collections of fluids. Codes of comparable breadth would be extremely lengthy to implement in the Jordan frame, and their fourth-order nature has consequences for speed and stability.

Considering the nature of the transformation for a perturbed model reveals an issue which to the best of our knowledge is as yet unappreciated. Perturbed systems in relativity exhibit the gauge issue -- the mapping between the fictitious background spacetime and the physical perturbed spacetime is non-unique. In cosmology, gauge freedoms allow one to eliminate four perturbative degrees of freedom (two scalar and two vector, where the classification is based on how the modes transform on a 3-surface), which specifies the slicing and threading we have chosen for our foliation. The transformation between the Einstein and the Jordan frames tangles this choice. We consider the impact of this and how one can resolve the resulting gauge ambiguities. We illustrate with a simple $f(R)$ model in a vacuum FLRW spacetime, but it should be emphasised that this issue will in principle occur in any perturbed spacetime and a wide range of extended theories of gravity.

We choose to work in the standard (``metric'') approach to cosmological perturbations (e.g. \cite{MukhanovFeldmanBrandenberger,MaAndBertschinger}) and leave potential issues in the 3+1 gauge-invariant and covariant (GIC) approach \cite{EllisBruni,Li:2007xn,Abebe:2011ry} for future study.\footnote{The issue of perturbed models of modified gravity in the 3+1 GIC approach has been considered in \cite{Carloni:2009gp}. The authors wish to thank Sergei Odintsov for bringing this work to our attention.} While many authors working on perturbed spacetimes in modified gravity choose to employ ``gauge-invariant'' variables (for example, \cite{Hwang1990CQG,MukhanovFeldmanBrandenberger}) or work in the Jordan frame (such as \cite{Hwang1996PRD,Song:2006ej,Carroll:2006jn,Tsujikawa:2007tg,delaCruzDombriz:2008cp,DeFelice:2010gb,Bertacca:2011wu}), the analysis of gauge problems is necessary for three reasons. Firstly, fixing a gauge (as in \cite{Hwang:1996np,Bean:2006up,Zhang:2007nk,Tsujikawa:2007gd}), is a valid approach, and any problems when transforming to the Einstein frame must be understood. Secondly, gauge-invariant variables are themselves based on variables in a particular gauge \cite{Malik:2008im}. The frequently-employed Bardeen potentials are, for example, gauge-invariant generalisations of the potentials in Newtonian gauge. If there is an issue with the choice of the ``natural'' gauge of a gauge-invariant perturbation, the perturbation itself cannot necessarily be trusted. Thirdly, our study is motivated not least by the use of Boltzmann codes in the Einstein frame. By far the widest used Boltzmann code is CAMB \cite{CAMB}, which is written in a frame equivalent to synchronous gauge.\footnote{Of the other modern Boltzmann codes, cmbeasy \cite{CMBEasy} can be used in synchronous or Newtonian gauges, and CLASS \cite{CLASS} is best tested in synchronous gauge although it was updated with support for Newtonian gauge after the release of the first version of this paper. Very recently, CosmoLib \cite{CosmoLib} was released, programmed entirely in Newtonian gauge. COMICS \cite{COSMICs} could be employed in both Newtonian and synchronous gauges, but was rapidly superseded by CMBFast \cite{CMBFast} which could only be used in synchronous gauge. Both of these codes are now obsolete, with CMBFast in turn superseded by CAMB and cmbeasy.  Our comments concerning CAMB and synchronous gauge equally apply to other Boltzmann codes in other gauges.} If there is an issue with the frame transformation and synchronous gauge, then the initial conditions for modifications of CAMB could be suspect, as could the final results even after being transformed back into the Jordan frame. Worse, if there is an issue the error is compounded: each transformation between frames will induce errors. Given that the use of CAMB (or similar codes) is of interest, the study of transformations of systems in synchronous gauge is forced upon us, despite its known shortcomings: synchronous gauge is not fully fixed \cite{Malik:2008im,Christopherson:2011ra} and ambiguities in the choice of threading remain.

The alternative formulation of $f(R)$ gravity employs the Palatini approach, where the metric and the connections are treated as independent objects. The Palatini approach is popular in cosmology (e.g. \cite{Capozziello:2010ef,Capozziello:2011et,Olmo:2011uz} and their references) and often employed in studies of cosmological perturbations (e.g. \cite{Koivisto:2005yc,Li:2006vi,Sotiriou:2008rp}). The nature of the transformation between the frames implies that the issues we highlight in this paper are also relevant in the Palatini approach. Since the field equations in the Palatini approach are intrinsically second-order the usefulness of the transformation is somewhat lessened, but if a system is more easily evaluated in the Einstein frame then the potential ambiguities should still be considered.

The gauge problem is discussed in \S\ref{Section:GaugeAmbiguity}A and illustrated in \S\ref{Section:GaugeAmbiguity}B. We close the paper in \S\ref{Section:Discussion} with a brief discussion.

We employ a signature $(-+++)$ and the Ricci tensor $R_{\mu\nu}=R^\alpha_{\phantom{\alpha}\mu\alpha\nu}=\Gamma^\alpha_{\mu\nu,\alpha}+\ldots$ (as in for example \cite{Gravitation,Maeda,DeFelice:2010aj}) and generally follow the notation of \cite{DeFelice:2010aj}. An overdot will represent differentiation with respect to conformal time $d\eta=dt/a$, and the conformal Hubble rate is written $\mathcal{H}=\dot{a}/a$.

\section{f(R) Gravity in the Einstein and Jordan Frames}
\label{Section:Model}
\subsection{f(R) Gravity in the Jordan Frame}
\noindent In $f(R)$ theory the Einstein-Hilbert Lagrangian density is replaced with a general function of the Ricci scalar, leading to the action
\be
\label{JFAction}
S=\frac{1}{2\kappa^2}\int d^4x\sqrt{-g}f(R)+\int d^4x\mathcal{L}_M\left(g_{\mu\nu},\Psi_M\right)
\ee
where $\Psi_M$ denotes the matter fields. The field equations follow by varying this action with respect to the metric,
\be
\label{JFEFE}
FR_{\mu\nu}-\frac{1}{2}fg_{\mu\nu}-\nabla_\mu\nabla_\nu F
+g_{\mu\nu}\nabla^\alpha\nabla_\alpha F=\kappa^2 T_{\mu\nu}^{(M)}, \quad
T_{\mu\nu}^{(M)}=-\frac{2}{\sqrt{-g}}\frac{\delta\mathcal{L}_M}{\delta g^{\mu\nu}} .
\ee
Here $F=\partial f/\partial R$. The matter stress-energy tensor obeys the standard conservation law,
\be
\nabla^\mu T_{\mu\nu}^{(M)}=0 .
\ee
This form of the theory is known as the ``Jordan Frame''. By introducing an auxiliary field it can be shown (e.g. \cite{Wands:1993uu,DeFelice:2010aj}) to be equivalent to a Brans-Dicke theory with $\omega_{\mathrm{BD}}=0$ so long as $\partial^2f/\partial R^2\neq 0$. In the Brans-Dicke representation the theory has also been described and studied as an example of ``extended quintessence'' (e.g. \cite{Pettorino:2008ez} and its references). The flat FLRW metric in the Jordan frame can be written
\be
ds^2=a^2(\eta)\left(-d\eta^2+\delta_{ij}dx^idx^j\right) .
\ee
For more details on $f(R)$ theory, see for example \cite{Capozziello:2007ec,Sotiriou:2008rp,DeFelice:2010aj} and their references.

\subsection{f(R) Gravity in the Einstein Frame}
\label{EinsteinFrameDefinition}
\noindent $f(R)$ gravity can be cast in a more familiar form through a conformal transformation
\be
g_{\mu\nu}\rightarrow \tg_{\mu\nu}=\Omega^2g_{\mu\nu} \Rightarrow \tg^{\mu\nu}=\Omega^{-2}g^{\mu\nu}, \quad \tg=\Omega^8g .
\ee
A conformal transformation leaves null geodesics invariant, implying that the causal structure of spacetime is unaffected by this mapping. Writing $\ln\Omega=\omega$, the Christoffel symbols, covariant derivatives of a scalar and a covector, the Ricci tensor, Ricci scalar, stress-energy tensor, stress-energy conservation transform \cite{Wald,Maeda,Faraoni:1998qx,Capozziello:2011et} as
\be
\label{Transformations}
\begin{array}{c}
\tilde{\Gamma}^\lambda_{\mu\nu}=\Gamma^\lambda_{\mu\nu}+\left(2\delta^\lambda_{(\mu}\nabla_{\nu)}\omega-g_{\mu\nu}\nabla^\lambda\omega\right), \quad\vspace{6pt}
\tNabla_\mu \phi=\nabla_\mu \phi, \quad\phantom{A_B^B} \tNabla_\mu v_\nu=\nabla_\mu v_\nu-\left(2\delta^\lambda_{(\mu}\nabla_{\nu)}\omega-g_{\mu\nu}\nabla^\lambda\omega\right)v_\lambda, \\
\tR_{\mu\nu}=R_{\mu\nu}-2\nabla_\mu\nabla_\nu\omega-g_{\mu\nu}\nabla^\alpha\nabla_\alpha\omega+2\nabla_\mu\omega\nabla_\nu\omega-2g_{\mu\nu}\nabla^\alpha\omega\nabla_\alpha\omega, \quad\vspace{8pt}
\tR=\Omega^{-2}\left(R-6\nabla^\mu\nabla_\mu\omega-6\nabla^\mu\omega\nabla_\mu\omega\right), \\
\tT_{\mu\nu}^{(M)}=\Omega^{-2}T^{(M)}_{\mu\nu}
\end{array}
\ee
and the matter is deflected from the new geodesics:
\be
\tNabla_\mu \tT_{(M)}^{\mu\nu}=-\tT_{(M)}\tNabla^\nu\omega .
\ee
Writing $f=RF+(f-RF)=RF-U$, the action (\ref{JFAction}) can then be rewritten as
\be
S=\int\sqrt{-\tg}\left[\frac{F}{2\kappa^2\Omega^2}\left(\tR+6\tNabla^\mu\tNabla_\mu\omega-6\tNabla^\mu\omega\tNabla_\mu\omega\right)
-\frac{U}{\Omega^4}\right]d^4x+\int\mathcal{L}_M\left(\Omega^{-2}\tg_{\mu\nu},\Psi_M\right)d^4x .\qquad
\ee
The Einstein frame is then defined by isolating a Ricci scalar in the action to serve as the Einstein-Hilbert Lagrangian density; this can be found by choosing
\be
\Omega^2=F
\ee
which requires the condition $F>0$ to be always satisfied. Casting the rest of the action into the form of a scalar field minimally-coupled to gravity,
\be
\kappa\tphi=-\frac{1}{2Q}\ln F, \qquad Q=-\frac{1}{\sqrt{6}},
\ee
produces
\be
S=\int\sqrt{-\tg}\left(\frac{1}{2\kappa^2}\tR-\frac{1}{2}\tNabla^\mu\tphi\tNabla_\mu\tphi
-\tV(\tphi)\right)d^4x
+ S_M\left(e^{2Q\kappa\tphi}\tg_{\mu\nu},\Psi_M\right) .
\ee
This action describes general relativity in the presence of a scalar field minimally-coupled to gravity but non-minimally coupled to matter -- an example of a ``coupled quintessence'' (e.g. \cite{Amendola:1999er,Amendola:1999dr,Amendola:2007yx,Pettorino:2008ez,Wintergerst:2009fh,Schrempp:2009kn}) model.\footnote{However, it should be noted that more recent coupled quintessence models such as that in \cite{Wintergerst:2009fh} tend to couple the scalar field only to particular species of matter, and do not easily admit a Jordan frame representation.} The stress-energy tensor and potential of the scalar field are
\be
\label{Potential}
\tT_{\mu\nu}^{(\phi)}=\tNabla_\mu\tphi\tNabla_\nu\tphi-\frac{1}{2}\tg_{\mu\nu}\tNabla^\sigma\tphi\tNabla_\sigma\tphi-\tV(\tphi)\tg_{\mu\nu}, \quad
\tV(\tphi)=\frac{RF-f}{2\kappa^2F^2}
\ee
and the field obeys the equation of motion
\be
\tNabla^\mu\tNabla_\mu\tphi-\frac{\partial\tV}{\partial\tphi}=-\kappa Q\tT^{(M)} .
\ee
For further details see for example \cite{DeFelice:2010aj}.

The FLRW metric in the Einstein frame is
\be
d\tilde{s}^2=\ta^2(\eta)\left(-d\eta^2+\delta_{ij}dx^idx^j\right)
\ee
where the coordinates are the same in the Einstein frame as in the Jordan frame -- this is possible because we are employing conformal instead of coordinate time.

\section{The Equivalence Between the Frames}
\label{Section:Equivalence}
\noindent The interpretation of this ``equivalence'' has a tangled history in the literature. Usage can be generally separated into two camps (see for example \cite{Faraoni:1998qx}, who identify a number of works to that date with one camp or the other, and \cite{Quiros:2011wb} which sets out clear definitions of ``equivalence''): it can be taken to imply that the physics in both frames is identical, or it can be taken to imply that a system set up in the Jordan frame can be solved in the Einstein frame as long as it is transformed back to the Jordan frame for interpretation. The former case relies on us clearly stating what ``physical equivalence'' means; in \cite{Deruelle:2010ht} (and, similarly \cite{Flanagan:2004bz}) the authors take the reasonable definition that the observables should remain the same, so long as the correct length and time-scales are employed. The latter case requires us to define the concept of the ``physical frame'', the one in which observations should be taken. This would be the frame in which it is natural to define our theory; if we are motivated, as in $f(R)$ gravity, by a modification of the Einstein-Hilbert action then the Jordan frame would be the physical frame. If instead we were motivated, as in coupled quintessence, by exotic couplings between a scalar field and the matter sector, then the Einstein frame would be the physical frame.

Particularly clear discussions of this issue are found in \cite{Faraoni:1998qx,Capozziello:2011et} and for some time it seems generally agreed that the ``equivalence'' is mathematical in nature, but since it occasionally reappears in the literature (see for instance \cite{Flanagan:2004bz,Nojiri:2006gh,Bhadra:2006rn,Capozziello:2005mj,Capozziello:2006dj,Faraoni:2006fx,Sotiriou:2007zu,Nozari:2009ds,Capozziello:2010sc,Deruelle:2010ht,Quiros:2011wb,Capozziello:2011wg} for some examples since 2004) we briefly re-address the question here.

``Physical equivalence'' between the frames implies that the general behaviour of solutions in the two frames would coincide. While the causal structure is naturally preserved by the conformal transformation, it is straightforward to find models with very different behaviours. For example, in polynomial gravity with $f(R)=\alpha R^n$ one can find \cite{Carloni:2004kp} a dust-dominated cosmology with
\be
a\propto \eta^{-2n/(2n-3)}, \quad
\ta\propto \eta^{(n-3)/(2n-3)}
\ee
in the Jordan and Einstein frames respectively. The deceleration parameter\footnote{This is defined with respect to the coordinate time $t$ as $q=-a(d^2a/dt^2)/\dot{a}^2$. In conformal time this becomes $q=-\dot{\mathcal{H}}/\mathcal{H}^2$. As such, $\tilde{q}$ is not the deceleration parameter $q$ transformed into the Einstein frame, but is the deceleration parameter one finds assuming the Einstein frame to be ``physical''. The discrepancy arises because, unlike the conformal time, coordinate time is transformed by the conformal mapping.} is then
\be
q=\frac{1}{2}\left(\frac{3-2n}{n}\right), \quad 
\tilde{q}=\frac{2n-3}{n-3} .
\ee
The universe in the Jordan frame is then accelerating if $n>3/2$, but is accelerating in the Einstein frame if $n\in (3/2,3)$. Choosing $n=4$ gives $q=-5/8$ and $\tilde{q}=5$. In the first case the universe is accelerating, while in the second it is decelerating. In this strict respect, the frames are clearly not physically equivalent. This example is similar to that in \cite{Capozziello:2010sc}; extreme forms are presented in \cite{Capozziello:2005mj,Capozziello:2006dj} where the authors find a phantom behaviour in one frame but a non-phantom frame in the other, and in \cite{Nozari:2009ds} where a model based on a non-minimally coupled scalar field exhibits a bouncing behaviour in the Jordan frame but not in the Einstein frame. An interesting example from outside cosmology is presented in \cite{Capozziello:2011wg}, which studies Birkhoff's theorem and argues that the Einstein frame representation is stable while the Jordan frame is not.

In the previous section the two frames are shown to be mathematically equivalent at the level of the action. It is quick to confirm that the field equations in the two frames are also equivalent to one another, in that a solution in the Jordan frame maps directly onto a solution in the Einstein frame and vice-versa. Consider the Einstein equations in the Einstein frame,
\be
\tR_{\mu\nu}-\frac{1}{2}\tg_{\mu\nu}\tR=\kappa^2\left(\tT^{(M)}_{\mu\nu}+\tT^{(\phi)}_{\mu\nu}\right) .
\ee
Employing the transformations (\ref{Transformations}) and the form of the scalar field stress-energy tensor (\ref{Potential}), it is straightforward to convert this to a Jordan frame equivalent,
\be
R_{\alpha\beta}-\frac{1}{2}g_{\alpha\beta}R-2\nabla_\alpha\nabla_\beta\omega+2g_{\alpha\beta}\nabla^\mu\nabla_\mu\omega
-4\nabla_\alpha\omega\nabla_\beta\omega+4g_{\alpha\beta}\nabla^\mu\omega\nabla_\mu\omega
=\kappa^2\Omega^{-2}T_{\alpha\beta}+\kappa^2\tV(\tphi)\Omega^2g_{\alpha\beta} .
\ee
Using now that $F=\Omega^2>0$ and $2\Omega^4\kappa^2\tV=\Omega^2R-f(R)$ produces the field equations in the Jordan frame (\ref{JFEFE}). Since the transformation is invertible, the converse naturally follows. Therefore
\be
\label{TransformationProof}
G_{\mu\nu}=\kappa^2T_{\mu\nu} \iff \tG_{\mu\nu}=\kappa^2\tT_{\mu\nu}
\ee
so long as $F>0$ is always obeyed. This implies that any solution of the field equations in the Jordan frame induces a solution in the Einstein frame, and vice-versa: the two frames are equivalent, even if only one frame is ``physical'' in the sense defined above. See for example \cite{Wald,Maeda,Faraoni:1998qx,Deruelle:2010ht} for other discussions and interpretations.

For our purposes it suffices to take the minimal view that the equivalence is mathematical and that observations should be made in the Jordan frame. This satisfies the interpretations of both camps; at worst, we have performed an unnecessary transformation before measuring the observables.

\section{Perturbations and Gauge Issues in $f(R)$ Cosmology}
\label{Section:GaugeAmbiguity}
\subsection{Transforming Perturbations between Frames: Gauge Ambiguities}
\noindent Applied to a linearly perturbed metric and writing $F$ for the background quantity, the conformal transformation between the frames is
\be
\tilde{g}_{\mu\nu}=\tilde{g}^{(0)}_{\mu\nu}+\tilde{h}_{\mu\nu}=(F+\delta F)\left(g^{(0)}_{\mu\nu}+h_{\mu\nu}\right) .
\ee
Assuming a flat FLRW background for illustration, all components of the metric imply
\be
\label{BackgroundTransformations}
\tilde{a}=\sqrt{F}a \Rightarrow \tH=\mathcal{H}+\frac{1}{2}\frac{\dot{F}}{F} .
\ee
The metric perturbation transforms as
\be
\label{PertConfTransform}
\tilde{h}_{\mu\nu}=Fh_{\mu\nu}+g_{\mu\nu}^{(0)}\delta F.
\ee
The perturbation can be expanded into scalar, vector and tensor modes \cite{Malik:2008im} by
\be
h_{\mu\nu}dx^\mu dx^\nu=a^2(\eta)\left(-2\Phi d\eta^2+\left(\partial_iB-S_i\right)d\eta dx^i+\left(-2\Psi\delta_{ij}+2\partial_i\partial_jE+2\partial_{(i}F_{j)}+h^{(T)}_{ij}\right)dx^idx^j\right)
\ee
with $\partial^iS_i=\partial^iF_j=\partial^ih_{ij}^{(T)}=\delta^{ij}h_{ij}^{(T)}=0$. Applying the transformation leads ultimately to
\be
\label{BardeenTransformations}
\tilde{\Phi}=\Phi+\dfrac{1}{2}\dfrac{\delta F}{F}, \quad
\tilde{\Psi}=\Psi-\dfrac{1}{2}\dfrac{\delta F}{F},
\ee
with all other metric quantities invariant. The transformation then induces both a lapse and a scalar 3-curvature regardless of the gauge chosen for $h_{\mu\nu}$. If for instance we employ uniform curvature gauge in the Jordan frame ($\Psi=E=0$), giving us two scalar degrees of freedom $\Phi$ and $B$, then upon transformation into the Einstein frame we apparently have \emph{three} scalar degrees of freedom: $\Phi$, $B$ and $\Psi$. Obviously these are not genuine degrees of freedom, being linear combinations of the two Jordan frame degrees of freedom. However, if the field equations are taken at face value, one should treat them as three. Erroneously employing the uniform curvature gauge field equations in the Einstein frame would in principle induce a gauge mode that would contaminate the results in the Einstein frame. Likewise, if we employed synchronous gauge in the Jordan frame, with $\Phi=B=0$ and the degrees of freedom $\Psi$ and $E$, upon transformation into the Einstein frame we would have three apparent degrees of freedom: $\Psi$, $E$ and the redundant lapse $\Phi$. Clearly if care is not taken the gauge is being tangled in the transformation between Jordan and Einstein frames -- only gauges possessing both a lapse and a curvature, such as Newtonian gauge, are unaffected.

The transformations of the matter and scalar field between frames provide a further constraint. Let $u^\mu u^\nu T^{(M)}_{\mu\nu}=\rho_M(1+\delta_M)$ and $u_{(M)}^\mu=(1/a)(1-\Phi,\partial^i v_{(M)})$, and expand the scalar field $\tpsi=\kappa\tphi$ as $\tpsi\rightarrow\tpsi+\delta\tpsi$. Then
\be
\label{FieldTransformations}
\trho_M=\frac{\rho_M}{F^2}, \quad \tdelta_M=\delta_M-2\frac{\delta F}{F}, \quad \tpsi=-\frac{1}{2Q}\ln F, \quad \delta\tpsi=-\frac{1}{2Q}\frac{\delta F}{F}
\ee
with the velocity remaining invariant. The density perturbation also changes, implying that if we chose uniform density gauge in the Jordan frame we would no longer be in uniform density gauge following the conformal transformation.

Note that this issue only affects scalar perturbations and not vectors or tensors. It is also important to note that this is not unique to cosmological models; any system which is separated into a background metric plus perturbations will be at risk.

The cause of this is that the group associated with the conformal transformation is more tightly constrained than that of GR. The gauges that lie within this group are those with $\Phi\neq 0$, $\Psi\neq 0$ and $\delta\rho\neq 0$; synchronous, uniform curvature (``flat'') and uniform density gauges do not belong to this group, while Newtonian (``longitudinal'' or ``Poisson'') and forms of comoving gauges do. Practically, there are three straightforward approaches to dealing with the problem:
\begin{itemize}
\item One could choose to work directly with the additional redundant degree(s) of freedom in the Einstein frame, keeping those terms in the field equations.
\item One could ``refix'' the gauge after a transformation. Transforming uniform curvature gauge between the frames induces a lapse; applying a gauge transform into Newtonian or back into uniform curvature gauge will reabsorb these, explicitly leaving a system with two degrees of freedom.
\item More physically, one could work in a gauge lying within the restricted group. The most obvious example is Newtonian gauge, while other options would include a subset of the comoving gauges with non-vanishing lapse and curvature.
\end{itemize}

It must be emphasised here that this is only an issue when the conformal transformation is applied -- while in either frame, one may choose to work with any gauge, as normal. For instance, one should be able to transform in Newtonian gauge into the Einstein frame, and then convert to uniform curvature gauge to undertake a calculation, before transforming back to Newtonian gauge and transferring the result into the Jordan frame. As a result studies in Jordan frame that employ a gauge outside of the restricted class (such as \cite{Bertacca:2011wu}) remain entirely valid.

While employing a system with apparent redundant degrees of freedom might seem a bit odd, it can be easily justified. By the equivalence of the field equations (\ref{TransformationProof}) any solution valid in the Jordan frame is equally valid in the Einstein frame. A corollary is that, \emph{as long as one employs the correct Einstein equations in the Einstein frame}, a Jordan frame solution will always be valid. However, these correct Einstein equations are not necessarily those that would be expected, and must additionally include the additional terms induced by the transformation. For example, the gauge-unfixed Hamiltonian constraint \cite{Malik:2008im} in the Einstein frame is
\be
3\tH\left(\dot{\tPsi}+\tH\tPhi\right)-\partial^i\partial_i\left(\tPsi+\tH\left(\dot{\tE}-\tB\right)\right)=-\frac{1}{2}\kappa^2\ta^2\delta\trho .
\ee
If we employ uniform curvature gauge in the Jordan frame, with $\Psi=E=0$, then after transforming into the Einstein frame we would na\"ively evolve the system using
\be
3\tH^2\tPhi+\tH\partial^i\partial_i\tB=-\frac{1}{2}\kappa^2\ta^2\delta\trho .
\ee
However, while we still have $\tilde{E}=0$, equation (\ref{BardeenTransformations}) implies that $\tilde{\Psi}\neq 0$. The correct version of the Hamiltonian constraint is then
\be
3\tH\left(\dot{\tPsi}+\tH\tPhi\right)-\partial^i\partial_i\left(\tPsi-\tH\tB\right)=-\frac{1}{2}\kappa^2\ta^2\delta\tilde{\rho} .
\ee
The validity of this system is guaranteed by equation (\ref{TransformationProof}).

This might be unpalatable on practical grounds: a great advantage of working in the Einstein frame is to leverage existing codes which are written in a set gauge, and evolving the extra degree of freedom will involve additional coding and testing. 
One does, however, remain free to apply a gauge transformation into a more standard gauge. It is this which we refer to as ``refixing''. The validity of the redundant system in the Einstein frame equally guarantees the validity of the refixed system given a well-defined gauge transformation. As such, gauges such as uniform curvature and uniform density gauges can be seen as being clean when transforming between the Jordan and Einstein frames -- and indeed are, so long as either the redundant system is evolved, or is reabsorbed following a gauge transformation. Failure to take one of these two steps will result in errors.

At this point it should again be emphasised that CMB codes are commonly written in synchronous gauge \cite{COSMICs,CMBFast,CAMB,CLASS}. The transformation into synchronous gauge contains an arbitrary function of space \cite{Malik:2008im}. While in conformal synchronous gauge the ambiguity can be removed by locking synchronous gauge to the velocity of cold dark matter, this additional step must be taken if one is intending to use such a code in the Einstein frame. If one takes the Jordan frame to be the physical frame (in which the Hubble rate, matter densities and magnitude of perturbations are set), then the initial conditions must be set up in the Jordan frame and then transferred into the Einstein frame. To employ these in a Boltzmann code a further transformation into synchronous gauge is required \emph{regardless of the gauge chosen in the Jordan frame}, and must be treated with the appropriate care.\footnote{An interesting point that can be made here is that the conformal transformation does not necessarily preserve adiabaticity of initial conditions -- that is, even if one sets adiabatic initial conditions in the Jordan frame, the introduction of an effective scalar field in the Einstein frame may induce a level of isocurvature, which should be taken into account. This issue is left to future study. The authors are grateful to David Mota for highlighting this point.}

The final option is to work in a gauge lying in the restricted group. This group requires the existence of a lapse, a spatial curvature and matter perturbations. In the Jordan frame we have four metric and two matter scalar degrees of freedom, $\Phi$, $B$, $\Psi$, $E$, $\delta\rho$ and $v$, of which we can remove two with gauge transformations. Of these, $\Phi$, $\Psi$ and $\delta\rho$ are non-zero in a gauge belonging to the restricted group, leaving us with the freedom to remove $B$, $E$ and $v$. In vacuo, this limits us to Newtonian gauge alone, while for systems with matter sources we have greater freedom and can also select comoving gauges with $v=0$ and either $E\neq 0$ or $B\neq 0$.

In the remainder of this paper we illustrate this issue with a simple example system.

\subsection{Gauge Issues in Vacuum Polynomial Gravity}
In this section we focus on polynomial gravity, with
\be
f(R)=\alpha R^n .
\ee
The background cosmology of this theory has been considered in detail in \cite{Carloni:2004kp}. These authors found the exact background solution both in vacuum and in the presence of matter fields,
\be
\label{aFuncT}
a\propto t^{(1-n)(2n-1)/(n-2)} .
\ee
The exponent diverges when $n\rightarrow 2$, suggesting that it is a particularly interesting case.

The situation is clearer in the Einstein frame. The scalar field potential (\ref{Potential}) becomes
\be
\kappa^2\tV(\tphi)=\frac{1}{2\alpha^{1/(n-1)}}\frac{n-1}{n^{n/(n-1)}}\exp\left(2\frac{n-2}{n-1}Q\kappa\tphi\right)
\ee
which is constant for $n=2$. For particular initial data or at late times we would then expect $(d\tphi/dt)^2\ll\tV(\tphi)$, and the spacetime in the Einstein frame will become de Sitter. For a field with $\dot{\tpsi}=0$, equation (\ref{BackgroundTransformations}) implies that the scale factor retains its character -- an exponential expansion in the Einstein frame will map onto an exponential expansion in the Jordan frame. When $n=2$, the solution in the Jordan frame is therefore de Sitter. This has been noted before in \cite{Capozziello:2003tk}, where $n=2$ is shown to correspond to $w_{\rm eff}=-1$, and in \cite{Carloni:2004kp} where it is stated that $\alpha\propto(n-2)=0$ corresponds to de Sitter space.

We take $n=2$ and for simplicity we work on superhorizon scales in vacuum and employ conformal time to ensure that our time coordinate remains invariant under the conformal transformation. This simplifies the treatment of the perturbations considerably.

\subsubsection{Background}
The Einstein field equations in the Jordan frame are
\be
\label{JFBackgroundEFF}
3\left(\frac{\dot{a}}{a}\right)^2+\frac{a^2}{4}R-3\frac{\ddot{a}}{a}+3\frac{\dot{a}}{a}\frac{\dot{R}}{R}=0, \quad
\frac{\ddot{a}}{a}+\left(\frac{\dot{a}}{a}\right)^2-\frac{\ddot{R}}{R}-\frac{\dot{a}}{a}\frac{\dot{R}}{R}-\frac{a^2}{4}R=0 .
\ee
The solution (\ref{aFuncT}) is in coordinate time. A solution for this system in conformal time is
\be
\label{JFa}
a(\eta)=(1+H_0(\eta_0-\eta))^{-1}, \quad \mathcal{H}=aH_0
\ee
where we have normalised such that $a(\eta_0)=1$. Conversely, writing the scalar field and energy density in units of $\kappa$, $\tpsi=\kappa\tphi$ and $\tR_\phi=\kappa^2\trho_\phi$, the Friedmann and continuity equations in the Einstein frame are
\be
3\left(\frac{\dot{\ta}}{\ta}\right)^2=\ta^2\tR_\phi, \quad
\dot{\tR}_\phi+3\tH\left(\tR_\phi+\tP_\phi\right)=0
\ee
where
\be
\tR_\phi=\frac{1}{2\ta^2}\dot{\tpsi}^2+\kappa^2\tV(\tphi), \quad \tP_\phi=\tR_\phi-2\kappa^2\tV(\tphi), \quad \tV(\tphi)=\frac{1}{8\alpha} .
\ee
The solution (\ref{JFa}) transformed into the Einstein frame is
\be
\ta=\dfrac{\sqrt{24\alpha}H_0}{1+H_0(\eta_0-\eta)}, \quad \tpsi=-\dfrac{1}{2Q}\ln\left(24\alpha H_0^2\right), \quad \tilde{w}_\phi=-1 .
\ee
The effective scalar field in the Einstein frame is frozen on a constant potential; as in the Jordan frame, this is de Sitter space.

This solution corresponds to the case where $\psi\equiv\mathrm{const.}$, which reduces the effective scalar field energy and pressure to constants. The system in the Einstein frame admits a more general case, corresponding to a field which is initially rolling in the potential before slowing through Hubble friction. The continuity equation for the scalar field can be written
\be
\frac{\partial}{\partial\ta}\left(\ta^6\tR_\phi\right)=\frac{3\ta^5}{4\alpha}
\ee
which can be solved to yield
\be
\tR_\phi=\frac{A}{\ta^6}+\frac{1}{8\alpha} .
\ee
If the initial conditions are tuned such that $A=0$, or when $\ta\gg 1$, then $\psi\equiv\mathrm{const}$ and this solution is the de Sitter limit found above. There is also a ``dynamic limit'' where the density decays rapidly with time, $\tR_\phi=A/\ta^6$, with $\ta=(2\sqrt{A/3}(\eta_0-\eta))^{1/2}$, which holds when $\ta\ll (8\alpha A)^{1/6}$. An implicit general solution for $a$ is
\be
\eta=\frac{4}{9}\left(\frac{16\alpha^2}{A}\right)^{1/6}\frac{\sqrt{3}\pi^2}{\left(\Gamma(2/3)\right)^3}-\frac{\sqrt{24\alpha}}{\ta}\;{}_2\mathcal{F}_1\left(\frac{1}{6},\frac{1}{2};\frac{7}{6};-\frac{8A\alpha}{\ta^6}\right)
\ee
where ${}_2\mathcal{F}_1({a};{b};x)$ is a regular hypergeometric function. In the limit $a\rightarrow\infty$ this tends to de Sitter space. (Note that this solution is not normalised such that $a(\eta_0)=1$.)

For our purposes it is sufficient to work in the de Sitter limit.

\subsubsection{Linear Perturbations in the Restricted Group}
In this section we find solutions for the perturbed vacuum spacetime in Newtonian gauge and demonstrate that the system is trivially valid in the Einstein frame in both the Newtonian and the uniform curvature gauges. Newtonian gauge lies in the restricted class of gauges, while uniform curvature gauge does not. The restricted gauge group consists of the usual gauge group in relativity with the additional conditions $\Phi\neq 0$ and $\Psi\neq 0$. The uniform density and uniform field gauges are other examples of gauges satisfying these conditions. On large scales where spatial derivatives can be neglected the vacuum perturbation equations \cite{Bean:2006up} are
\be
\begin{array}{c}
F\left(\Phi-\Psi\right)+\dfrac{\partial F}{\partial R}\delta R=0, \vspace{8pt} \\
6\mathcal{H}F\left(\dot{\Psi}+\mathcal{H}\Phi\right)+3\dfrac{\partial F}{\partial R}\dot{\mathcal{H}}\delta R
-3\mathcal{H}\dfrac{\partial}{\partial\eta}\left(\dfrac{\partial F}{\partial R}\right)\delta R
-3\mathcal{H}\dfrac{\partial F}{\partial R}\delta\dot{R}+3\dot{F}\left(2\mathcal{H}\Phi+\dot{\Psi}\right)=0
\end{array}
\ee
where it is understood that $F$ and $R$ are background quantities. Inserting our model with $F=2\alpha R$ and using
\be
\delta R=-\frac{6}{a^2}\left(2\frac{\ddot{a}}{a}\Phi+\mathcal{H}\dot{\Phi}+3\mathcal{H}\dot{\Psi}+\ddot{\Psi}\right)
\ee
one can find that
\be
\label{JFSolutions}
\Psi=\Psi_1+\frac{\Psi_2}{a}+\frac{\Psi_3}{a^3}, \quad \Phi=\frac{\Psi_2}{a}-\frac{\Psi_3}{a^3}-\Psi_1
\ee
is a general solution. Therefore
\be
\label{dFoF}
\frac{\delta F}{F}=2\left(\Psi_1+\frac{\Psi_3}{a^3}\right) .
\ee

This solution is transformed to the Einstein frame using the transformations defined in equations (\ref{BardeenTransformations}) and (\ref{FieldTransformations}), giving $\tPhi=\tPsi$ and
\be
\label{EFN}
\tPhi=\dfrac{\sqrt{24\alpha}H_0\Psi_2}{\ta}, \quad
\delta\psi=-\dfrac{1}{Q}\left(\Psi_1+\dfrac{(24\alpha)^{3/2}H_0^3\Psi_3}{\ta^3}\right) .
\ee
Then $\dot{\psi}=0$ at a background level, implying $8\alpha\tR_\phi=1$ and $\delta\tR_\phi=\delta\tilde{P}_\phi=0$. The Newtonian potentials are equal, as should be expected since in the Einstein frame, unlike the Jordan frame, the anisotropic stress vanishes. The Klein-Gordon equation, Hamiltonian constraint and shear evolution equations \cite{Malik:2008im} in this frame are
\be
\tH\left(\dot{\tPhi}+\tH\tPhi\right)=0, \quad \tPhi=\tPsi, \quad \delta\ddot{\tpsi}+2\tH\delta\dot{\tpsi}=0
\ee
which are clearly satisfied. This illustrates the expected equivalence between the frames: a system perturbed in Newtonian gauge can be evolved in either the Jordan or the Einstein frames, to yield a consistent result in the Jordan frame. The same holds for any gauge that lies within the restricted group.

A scalar quantity transforms under a gauge transformation $\xi^\mu=(\alpha,\partial^i\beta)$ as $\delta A_{\rm F}=\delta A+\dot{A}\alpha$, and so is gauge-invariant at linear order if it is constant on the background. As a result, the field perturbation is gauge-invariant because $\dot{\tpsi}=0$, and will therefore always satisfy its equation of motion. The transformation from Newtonian into uniform curvature gauge, defined by choosing $\Psi_{\rm F}=E_{\rm F}=0$ \cite{Malik:2008im}, is
\be
\tPhi_{\rm F}=\tPhi+\tPsi+\frac{\partial}{\partial\eta}\left(\frac{\tPsi}{\tH}\right), \quad \tB_{\rm F}=-\frac{\tPsi}{\tH},
\ee
where a subscript F denotes uniform curvature gauge. Uniform curvature gauge was used in, for example, \cite{Hwang:1996bc}. The lapse and shift in uniform curvature gauge are then
\be
\label{EFFlat}
\tPhi_{\rm F}=0, \quad \tB_{\rm F}=-\frac{24\alpha H_0^2}{\ta^2}\Psi_2
\ee
The Hamiltonian constraint and shear evolution become
\be
\label{FlatGaugeEFE}
3\tH^2\tPhi_{\rm F}=0, \quad
\dot{\tB}_F+2\tH\tB_F=0
\ee
which are clearly satisfied.

\subsubsection{Refixing a Gauge}
Consider the system in the uniform curvature gauge in the Jordan frame. Transforming the Jordan frame solutions (\ref{JFSolutions}) into uniform curvature gauge gives
\be
\label{JFFlat}
\Phi_{\rm F}=-\Psi_1-4\frac{\Psi_3}{a^3}, \quad B_{\rm F}=-\frac{1}{aH_0}\left(\Psi_1+\frac{\Psi_2}{a}+\frac{\Psi_3}{a^3}\right) .
\ee
Since the background Ricci scalar is constant, the perturbed Ricci scalar is gauge-invariant and $\delta F/F$ is given by equation (\ref{dFoF}). Transforming into the Einstein frame gives
\be
\label{EinsteinFrameFlat}
\begin{array}{c}
\tPhi_{\rm F,2}=-3(24\alpha)^{3/2}H_0^3\dfrac{\Psi_3}{\ta^3}, \quad
\tPsi_{\rm F,2}=-\Psi_1-(24\alpha)^{3/2}H_0^3\dfrac{\Psi_3}{\ta^3}, \vspace{8pt}\\
\tB_{\rm F,2}=-\dfrac{\sqrt{24\alpha}}{\ta}\left(\Psi_1+\sqrt{24\alpha}H_0\dfrac{\Psi_2}{\ta}+(24\alpha)^{3/2}H_0^3\dfrac{\Psi_3}{\ta^3}\right),
\end{array}
\ee
which evidently differ from (\ref{EFFlat}). In particular, since the lapse is non-vanishing the Hamiltonian constraint is satisfied only if $\Psi_1=0$, while the stress evolution equation is only valid if $\Psi_1=\Psi_3=0$ -- neither of which are acceptable.

To illustrate, consider a model in units where $\alpha=H_0=1$, and set the initial conditions $\Psi=1$, $\dot{\Psi}=0$, $\ddot{\Psi}=1/2$ in the Jordan frame at $a=1$. Then we find
\be
\Psi_1=\frac{7}{6}, \quad \Psi_2=-\frac{1}{4}, \quad \Psi_3=\frac{1}{12} .
\ee
Transferring into the Einstein frame, we find that $a=1$ corresponds to $\ta=\sqrt{24}$. On this time-slice equation\newline (\ref{EinsteinFrameFlat}) implies that $\tB_{F,2}(\ta=\sqrt{24})=-1$. Equations (\ref{FlatGaugeEFE}) then implies that
\be
\tB_{F,2}=-\frac{24}{\ta^2}
\ee
is the ``solution'' in the Einstein frame. However, equation (\ref{EFFlat}) shows that
\be
\tB_{F}=-\frac{24}{\ta^2}\Psi_2=\frac{6}{\ta^2} .
\ee
The discrepency is of the same order of magnitude as the perturbation itself. 
Even worse, transferring the lapse gives $\tPhi_{\rm F,2}(\ta=\sqrt{24})=-1/4$ on the initial hypersurface but equations (\ref{FlatGaugeEFE}) enforce $\tPhi_{\rm F}=0$ at all other times.

This illustrates the main result of our paper: if you na\"ively transfer a system in an unsafe gauge between the Jordan and Einstein frames, you will inevitably introduce errors of the same order of magnitude as the perturbations themselves.

In the previous section we argued two approaches to dealing with this: refixing the gauge, and employing a redundant set of field equations. Consider first refixing the gauge to uniform curvature gauge by reabsorbing the spatial curvature with the general transformations
\be
\tPhi_{\rm RF,F}=\tPhi_{\rm F,2}+\tPsi_{\rm F,2}+\frac{\partial}{\partial\eta}\left(\frac{\tPsi_{\rm F,2}}{\tH}\right), \quad \tB_{\rm RF,F}=\tB_{\rm F,2}-\frac{\tPsi_{\rm F,2}}{\tH}
\ee
implying $\tPhi_{\rm RF,F}=0$ and $\tB_{\rm RF,F}=-24\alpha H_0\Psi_2/\ta^2$ in agreement with (\ref{EFFlat}). Likewise, refixing to Newtonian gauge,
\be
\tPhi_{\rm RF,N}=\tPhi_{\rm F,2}+\dot{\tB}_{\rm F,2}+\tH\tB_{\rm F,2}=\sqrt{24\alpha}H_0\frac{\Psi_2}{\ta}, \quad \tPsi_{\rm RF,N}=\tPsi_{\rm F,2}-\tH\tB_{\rm F,2}=\tPhi_{\rm RF,N}
\ee
in agreement with (\ref{EFN}). These illustrate the validity of refixing the gauge to reabsorb the unexpected degree of freedom.

We can also consider the unfixed system in the Einstein frame. We have three scalar degrees of freedom: $\tPhi_{\rm F}$, $\tPsi_{\rm F}$ and $\tB_{\rm F}$. Retaining these three variables, the vacuum field equations on large scales  \cite{Malik:2008im} are
\be
3\tH\left(\dot{\tPsi}_{\rm F,2}+\tH\tPhi_{\rm F,2}\right)=0, \quad
\dot{\tB}_{\rm F,2}+2\tH\tB_{\rm F,2}+\tPhi_{\rm F,2}-\tPsi_{\rm F,2}=0, \quad
\ddot{\tPsi}_{\rm F,2}+2\tH\dot{\tPsi}_{\rm F,2}+\tH\dot{\tPhi}_{\rm F,2}+\left(2\dot{\tH}+\tH^2\right)\tPhi_{\rm F,2}=0
\ee
noting that the shear $\tilde{\sigma}=\dot{\tilde{E}}-\tB$. These field equations are also clearly satisfied by equations (\ref{EinsteinFrameFlat}).

\section{Discussion}
\label{Section:Discussion}
In this paper we have discussed a technical subtlety of the conformal transformation between the Jordan and Einstein frames that to the best of our knowledge has not been highlighted before: na\"ively transferring a perturbed system between the frames tangles your choice of gauge.  It should be emphasised that this effect is \emph{calculational}, not physical, and that if sufficient care has been taken no previous results are changed by this. It is also extremely important to note that this issue is not restricted to $f(R)$ gravity or to the study of cosmological perturbations, and that it potentially arises whenever a perturbed system is conformally transformed regardless of the background metric or the model of gravity. $f(R)$ theory and vacuum cosmology provide a useful, straightforward illustration. While we have concretely demonstrated the issue only for this vacuum $f(R)$ system, in principle it arises generically, although demonstrating this explicitly is beyond the scope of this paper.

The properties of the conformal transformation restrict us to a particular set of gauges which we refer to as the ``restricted group'': gauges that possess a lapse, a spatial curvature, and a density perturbation. This restriction only applies when the transformation is performed. At all other times one is free to work in any convenient gauge. However, we additionally argued that there are alternatives to working in the restricted group -- one can instead choose to ``refix'' the gauge, or to evolve the system using the redundant Einstein field equations.

We have illustrated our arguments with a simple vacuum system, showing that there are no ambiguities introduced by equations (\ref{BardeenTransformations}, \ref{FieldTransformations}), so long as one takes such care. If this is not done and the results of a transformation from a gauge outside the restricted group are interpreted as if they themselves lie in that gauge, the system will be inaccurately specified.

Perhaps the most important place where this issue will arise is in the treatment of initial conditions. For instance, Boltzmann codes are frequently \cite{COSMICs,CMBFast,CAMB,CLASS} written in synchronous gauge, which lies outside of the restricted group. Interpreting the Jordan frame as physical, the initial conditions must be set in the Jordan frame. However, regardless of the gauge chosen, the transformation into the Einstein frame induces a spurious lapse function which must be reabsorbed if the results of the calculation are to make any sense. This itself introduces an additional problem: the transformation to synchronous gauge is not fully specified, and introduces an arbitrary function of the 3-coordinates which must itself be removed with care. The alternative is to include the redundant degrees of freedom in the Boltzmann codes, which requires much additional effort. A careless study without attention to these issues would either leave the system with the spurious lapse, or risk introducing a gauge mode. Some Boltzmann codes \cite{COSMICs,CMBEasy,CLASS} include modules written in Newtonian gauge, which lies within the restricted group, and studies of extended gravity that employ these codes remain consistent. However, COSMICS is very outdated, while cmbeasy and CLASS are programmed in C++ and C respectively; given the widespread use of Fortran in cosmology, its large codebase and the extensive testing it has undergone, CAMB remains the dominant Boltzmann code and the study of synchronous gauge is necessary.\footnote{CosmoLib \cite{CosmoLib} is also written in Fortran but in Newtonian gauge; however it has appeared too recently to yet have much of an impact. However, it could prove extremely useful for studies of modified gravity.} Similar arguments apply to the initial conditions employed in n-body simulations of modified gravity performed in the Einstein frame.

The other time at which the gauge ambiguity becomes important is at the end of a calculation. Taking the Jordan frame to be physical, the results of a simulation in the Einstein frame must be transformed back. Given that calculations are frequently performed in synchronous gauge then one must either transform into a safer gauge before performing the conformal transformation, or deal with the redundant system in the Jordan frame. Otherwise one runs a serious risk of contamination. At this stage it seems easier to refix the gauge in the Jordan frame to a fully-specified gauge (such as Newtonian or uniform curvature), or to leave the system unspecified. So long as one does so consistently, observables will not be affected, since gauge transformations cannot change an observation.

It is interesting to note that the lensing potential $\Phi+\Psi$ is uncontaminated by the transformation. As such, if one wishes to calculate the weak lensing signal on a constant-time hypersurface, this can be achieved transforming from the Jordan frame into the Einstein frame in any gauge and not worrying about refixing the system \cite{Schimd:2004nq}. For a static (or quasi-static) system, this will be a good approximation. However, carelessly evolving the system without due care will in principle introduce errors, so that the frame invariance does not imply such a straightforward estimate of the integrated Sachs-Wolfe effect.

In summary, we recommend that authors exploiting the Einstein frame to study extended theories of gravity work in the restricted group. This removes the need for additional gauge transformations to reabsorb the extra apparent degrees of freedom, or the need to employ redundant Einstein equations. If one chooses the latter options, care should be taken to ensure that the result remains consistent.

\acknowledgments{The authors wish to thank David Mota, Sanjeev Seahra, Mikjel Thorsrud and Hans Winther for extremely useful discussions.}

\bibliography{fRBib}

\end{document}